\documentclass[journal]{IEEEtran}
\IEEEoverridecommandlockouts
\usepackage{cite}
\usepackage{amsmath,amssymb,amsfonts}
\usepackage{algorithmic}
\usepackage{graphicx}
\usepackage{textcomp}
\usepackage{xcolor}
\usepackage{lettrine}
\def\BibTeX{{\rm B\kern-.05em{\sc i\kern-.025em b}\kern-.08em
    T\kern-.1667em\lower.7ex\hbox{E}\kern-.125emX}}

\usepackage[hidelinks]{hyperref}

\begin{document}

\title{Distributed maze exploration using multiple agents and optimal goal assignment\\
\thanks{Research was possible due to the funding from the European Union’s Horizon Europe research and innovation programme under grant agreement No 101073876 (Ceasefire).}}

\author{Manousos Linardakis\thanks{Manousos Linardakis is with the Department of Informatics and Telematics, Harokopio University of Athens, Athens, Greece (e-mail: manouslinard@gmail.com, it22064@hua.gr). ORCID: 0009-0001-7060-848X.}, Iraklis Varlamis\thanks{Iraklis Varlamis is with the Department of Informatics and Telematics, Harokopio University of Athens, Athens, Greece (e-mail: varlamis@hua.gr). ORCID: 0000-0002-0876-8167.}, Georgios Th. Papadopoulos\thanks{Georgios Th. Papadopoulos is with the Department of Informatics and Telematics, Harokopio University of Athens, Athens, Greece (e-mail: g.th.papadopoulos@hua.gr). ORCID: 0000-0003-1686-421X.}
}

\maketitle

\begin{abstract}
Robotic exploration has long captivated researchers aiming to map complex environments efficiently. Techniques such as potential fields and frontier exploration have traditionally been employed in this pursuit, primarily focusing on solitary agents. Recent advancements have shifted towards optimizing exploration efficiency through multiagent systems. However, many existing approaches overlook critical real-world factors, such as broadcast range limitations, communication costs, and coverage overlap. This paper addresses these gaps by proposing a distributed maze exploration strategy (CU-LVP) that assumes constrained broadcast ranges and utilizes Voronoi diagrams for better area partitioning. By adapting traditional multiagent methods to distributed environments with limited broadcast ranges, this study evaluates their performance across diverse maze topologies, demonstrating the efficacy and practical applicability of the proposed method. The code and experimental results supporting this study are available in the following repository: \href{https://github.com/manouslinard/multiagent-exploration/}{https://github.com/manouslinard/multiagent-exploration/}.
\end{abstract}


\section{Introduction}
\label{sec:introduction}
\lettrine{E}{xploration} of mazes has interested researchers in the recent years, offering a window into the cognitive capabilities of various organisms, such as mice \cite{mice_maze}, and more recently, the artificial intelligence capabilities of robotic systems \cite{robotics_maze2, robotics_maze1}. While algorithms specifically designed for single-agent maze exploration are widely utilized \cite{singlerobot1, singlerobot2}, the adaptation of these algorithms to multiagent systems \cite{dorri2018multi, multi_comm3, multiagent_ai} presents persistent challenges, underscoring the necessity for novel solutions.

The task of mapping unknown mazes for exploration necessitates strategic methodologies. Robots traverse the maze, documenting their journey and observations to construct a map as they advance \cite{multiexplore_1, multiexplore_2}. By employing sensors, they identify walls and obstacles, enabling them to make informed decisions regarding their subsequent movements. Upon mapping a significant portion of the maze, robots strategize their routes towards their objectives, which could be reaching the exit or locating specific targets within the maze. This process involves analyzing the constructed map and selecting the most efficient path forward.

Using multiple agents to explore the maze expedites the mapping process and enhances the coverage of the resulting map. However, this approach introduces challenges, including the need for effective agent communication to share information and organize exploration to prevent collisions or obstructions in their paths \cite{multi_comm1, multi_comm2, multi_adversarial}. Additionally, multiagent exploration may lead to revisiting previously explored areas, extending total mapping time. Distributing the maze exploration task to multiple agents offers a solution to these problems, as robots can first explore their assigned regions individually, reducing redundant coverage, the number of collisions, and the need for extensive communication compared to exploring the entire maze map collectively.

There are several methods to partition a maze into sub-areas, including clustering algorithms such as K-means, which typically suffer from local optima and potentially lead to higher robot dispersion \cite{li2015dynamic}. Other approaches, like the Effective Regions of Movement (ERM) \cite{magsino2022homogeneous, magsino2022enhanced}, are available but require additional information (e.g., agent capacity, density, and speed), which is not easily acquired when exploring an unknown maze. Moreover, ERM is applied to known maps and is typically used for enhancing and understanding networks, such as those involving cars. In this paper, we employ Voronoi partitioning \cite{aurenhammer2000voronoi, voronoi_algo} centered around each agent, due to its suitability for unknown areas. Voronoi partitioning has been proven effective for segmenting 2D spaces, as demonstrated by Wu \textit{et al.} \cite{wu2007voronoi}, who compared it with K-means for area coverage. Their results showed that Voronoi partitioning was consistently faster than K-means across all examined maps.

Our proposed approach, named CU-LVP, is designed to address the distributed maze problem. Experiments detailed in Section \ref{sec:result_plots} showcase the method's efficiency, low computational complexity, minimal communication, and reduced repetitive coverage. We compare our method with other state-of-the-art techniques adapted to the distributed maze problem. To evaluate performance, we conduct extensive experimental evaluations across a variety of maze topologies, including 30x30 randomly generated mazes with varying obstacle densities. These experiments prove CU-LVP's efficacy and provide a comprehensive assessment of each method's performance. They offer valuable insights into their effectiveness in exploring distributed mazes and their practical applicability in real-world robotic systems. Additionally, by integrating constraints on broadcast ranges across all evaluated methods, we aim to emulate real-world communication limitations. By incorporating these constraints, we enhance the robustness of the exploration strategies, making them more applicable to real-world scenarios.

The main contributions of this paper can be summarized as follows:
\begin{itemize}
    \item A detailed review and classification of the recent exploration approaches using multiple agents.
    \item An introduction to the novel method CU-LVP, designed for covering distributed mazes efficiently. 
    \item An extensive implementation of established multiagent maze exploration techniques, applied in distributed mazes through the integration of Voronoi diagrams.
    \item A comparative evaluation of the algorithms, using various metrics that examine different aspects of the exploration task (e.g. time, distance, computational and communication cost).
\end{itemize}

Section \ref{sec:relwork} offers an overview of related work in state-of-the-art multiagent maze and area exploration. Section \ref{sec:proposed_method} details the proposed approach and Section \ref{sec:experimental_eval} explains the experimental evaluation process that we applied. Section \ref{sec:result_plots} examines the achieved results, whereas Section \ref{sec:conclusions} presents the main conclusions of this work and future work.

\section{Related Work} \label{sec:relwork}

\begin{figure*}[t]
  \centering
  \includegraphics[width=0.9\textwidth]{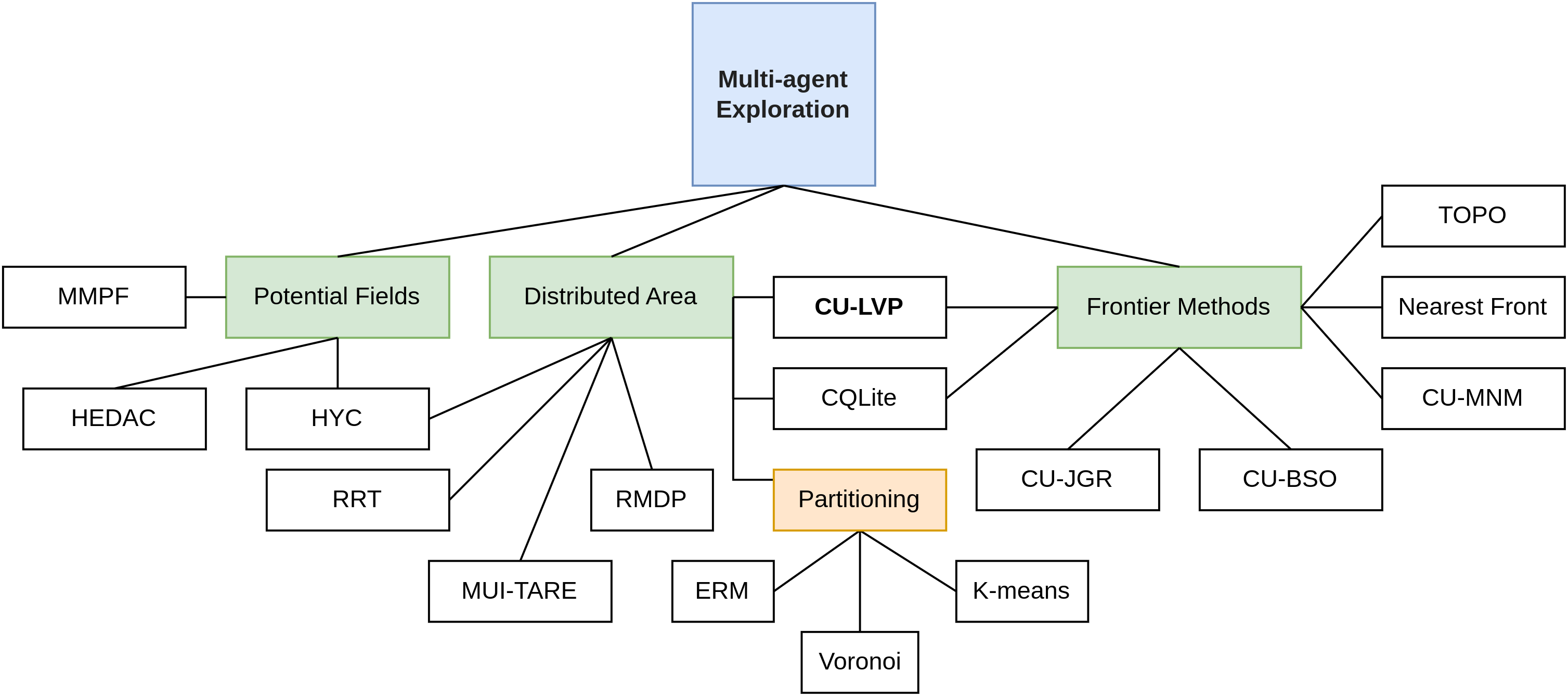}
  \caption{Hierarchical Structure of the Related Work (Section \ref{sec:relwork}). Method in \textbf{bold} (CU-LVP) is our proposed method.}
  \label{fig:rel_work_hierarchy}
\end{figure*}

Exploring the field of multiagent maze exploration encompasses a variety of innovative strategies and methodologies. This section categorizes and reviews significant contributions, highlighting the implementation details of each approach. We begin with distributed area exploration techniques, which emphasize on area/maze partition for efficient task allocation and coordination among agents. Next, we examine potential field methods, where agents use potential/temperature fields for maze coverage. Finally, we discuss frontier-based exploration, where the exploration process is based on current frontiers—the unexplored boundaries of the known environment. Fig. \ref{fig:rel_work_hierarchy} organizes the related work in a hierarchical structure for enhanced clarity and comparison.

\subsection{Distributed Area Methods}

Lozenguez \textit{et al.} \cite{lozenguez2011map} introduce a method for efficiently allocating exploration tasks among multiple mobile robots. In the context of this paper, we will refer to this method as RMDP. The approach uses a combination of Road-Map techniques and Markovian Decision Processes (MDPs) to manage the exploration of areas marked by points of interest. By representing spatial knowledge as a dynamically adjustable graph of waypoints and partitioning this graph into regions, the method enables optimal task assignments based on real-time evaluations. As part of the R-Discover project, this research supports the use of UAVs for initial mapping, followed by ground robots for detailed exploration. Robots operate under a hierarchical control system, with leaders dynamically reallocating tasks to maximize efficiency during the mission.

Yan \textit{et al.} \cite{yan2023mui} proposed MUI-TARE to address the intricate task of multiagent exploration in confined 3D spaces with unknown initial agent poses. Their work focuses on efficiently exploring these environments and effectively merging agents' sub-maps. Traditional strategies often struggle with merging sub-maps due to false overlap detections, leading to inaccuracies. To combat this, the authors propose a novel lidar-based approach that intelligently balances sub-map merging robustness and exploration efficiency. The approach allows agents to adaptively repeat each other's trajectories based on merging quality, minimizing redundant exploration. Additionally, it extends a single-agent hierarchical strategy to multiple agents, enhancing exploration efficiency. Experimental results demonstrate efficiency and exploration accuracy over baselines, ensuring robust sub-map merging in complex 3D environments.

Gui \textit{et al.} \cite{gui2023decentralized} tackled the challenge of efficient distributed exploration for UAVs, particularly in search and rescue scenarios. They present a novel cooperative strategy where UAVs dynamically explore distinct regions, enhancing efficiency while minimizing redundancy. Using a dynamic centroid-based approach, the 3D space is partitioned for each UAV, enabling independent target generation within their designated zones. They propose a next-best-view method employing a rapidly exploring random tree (RRT) to foster cooperative exploration and navigate unknown areas. Comparative evaluations against three classical methods underscore the superiority of their approach in both simulation and real-world experiments, marking a significant advancement in swarm UAV capabilities for dynamic environments like search and rescue missions.

Latif \textit{et al.} \cite{latif2024communication} introduced CQLite, a distributed Q-learning technique tailored to address the challenges of coordinating multiple mobile robots in the autonomous exploration of complex environments. Unlike conventional approaches reliant on internal global maps for navigation, CQLite prioritizes minimizing communication and information exchange costs between robots while ensuring swift convergence and coverage. By leveraging ad hoc map merging and selectively sharing updated Q-values at newly discovered frontiers, the method significantly reduces communication overhead. Theoretical analysis and numerical validation conducted on simulated indoor maps with multiple robots underscore the efficiency of the proposed approach.

\subsection{Potential Fields Methods}
Jincheng Yu \textit{et al.} \cite{yu2021smmr} introduced the Multirobot Multitarget Potential Field (MMPF), a novel approach that uses potential fields for maze exploration. The robots operate concurrently, utilizing a simultaneous localization and mapping (SLAM) technique for navigation and mapping tasks. Notably, MMPF does not initially implement any partitioning of the environment. Instead, the research emphasizes localization, dispersing agents across disparate areas of the environment. As agents subsequently explore overlapping regions, localization, and information exchange occur to facilitate comprehensive mapping of the area.

Crnkovic \textit{et al.} \cite{hedac_maze, hedac_og} proposed HEDAC, a robust algorithm designed for both area and maze exploration. By leveraging temperature fields to compute attractive forces, HEDAC excels in prioritizing the exploration of unknown regions while mitigating collisions. Specifically, it creates a strong attractive force for unexplored areas and a weaker force for obstacles and other agents. HEDAC has demonstrated efficacy in uncertain conditions, such as variations in sensor readings and unpredictable environmental fluctuations, as documented in \cite{ivic2020motion}.

A similar approach to HEDAC that utilizes Voronoi diagrams to divide an area into subregions is presented by Zheng \textit{et al.} \cite{zheng2022distributed}, which we will refer to as HYC in this paper. In HYC, each subregion is explored by an agent using a temperature field-induced control strategy. For our study, we have implemented HEDAC instead of HYC because, although both methods use temperature fields, HEDAC is specifically designed for maze exploration with many or fewer obstacles, whereas HYC
is not clearly implemented for such scenarios. The tested areas in the HYC paper are described as concave regions formed by removing small squares from larger ones, and the agent trajectories in the provided figures suggest that the tested areas are largely obstacle-free. Therefore, HEDAC has been deemed a more suitable choice for our task, given its effectiveness in both high and low-obstacle-density environments. Additionally, we have modified HEDAC to address the distributed maze problem, incorporating principles from the HYC paper, as detailed in Section \ref{subsec:baseline_methods}.

\subsection{Frontier Methods}
Exploring multiagent systems has led to a variety of methods and algorithms, with the nearest-frontier method being notable for its simplicity and effectiveness. Initially proposed by Yamauchi \cite{yamauchi1998frontier}, this method focuses on finding the shortest path to the closest frontier. Despite its efficiency, recent advancements in exploration strategies have introduced more sophisticated techniques, particularly those that use utility functions to enhance frontier selection. Cost-utility methods, for example, improve upon the deterministic nearest-frontier approach by incorporating additional utility functions, thus optimizing both exploration efficiency and decision-making processes.

Marjovi \textit{et al.} \cite{cu_mnm_paper} present a cost-utility approach, referred to here as CU-MNM, which effectively explores mazes while detecting fires. The utility function in CU-MNM is determined by the distance of the frontier from all robots. This decentralized approach promotes communication and collaborative decision-making among the robots.

Another cost-utility method, introduced by Julia \textit{et al.} \cite{cu_jgr}, uses the expected information gain in goal cells as the utility function to facilitate efficient exploration. For this paper, this method will be referred to as CU-JGR.

\begin{figure}
\centering
\includegraphics[width=0.3\textwidth]{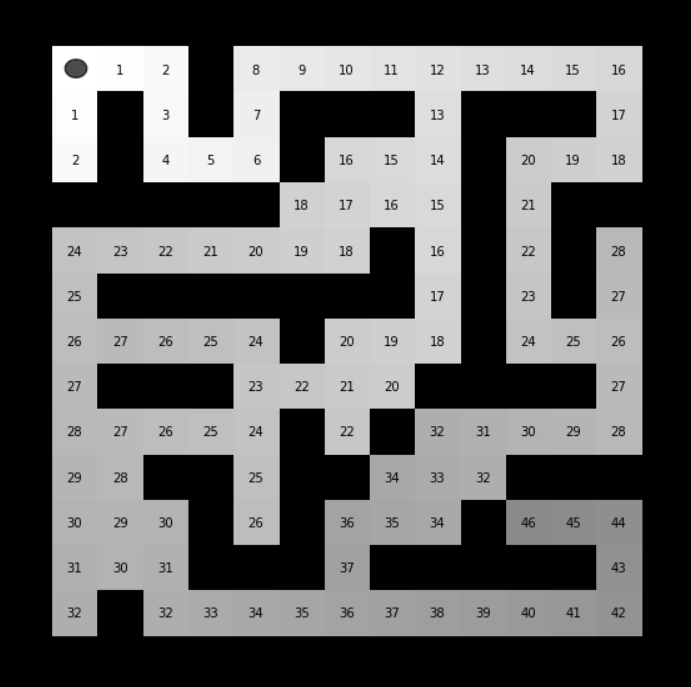}
\caption{Wavefront distance calculation for the entire maze. The gray dot indicates the agent's position, whereas each empty cell contains the distances from the agent. The black cells represent obstacles.}
\label{fig:wavefront}
\end{figure}

Bautin \textit{et al.} \cite{cu_bso} propose a computationally efficient frontier allocation method, referred to as CU-BSO in this paper, which encourages a balanced spatial distribution of robots within the environment. This method uses wavefront propagation \cite{wavefront}, originating from each frontier to quickly create a distance matrix (cost matrix) for all cells, and then selects frontiers with fewer nearby robots. The use of wavefront propagation underscores reduced communication requirements among robots. An example of the wavefront algorithm applied to a maze environment is shown in Fig. \ref{fig:wavefront}.

Beyond cost-utility functions, Zhang \textit{et al.} \cite{zhang2022mr} proposed the TOPO exploration method, designed to enhance multiagent exploration in communication-constrained environments. TOPO employs a frontier exploration strategy with a greedy approach, calculating the center of existing vertices whenever a new vertex is added. TOPO also constructs topological maps as robots navigate through the environment, reducing the data required for path planning and communication. Unlike traditional methods that depend on occupancy grid maps, TOPO significantly lowers data transfer requirements, as demonstrated in both simulated and real-world scenarios. When compared to the MMPF method \cite{yu2021smmr}, TOPO was shown to be faster in exploring the area through extensive experimentation.

Our approach (CU-LVP) contributes to the existing literature by introducing an efficient cost-utility method tailored to address the challenges of distributed maze coverage. Through rigorous experimentation, we have demonstrated that our method outperforms comparative approaches in terms of speed, achieving faster maze exploration while imposing minimal communication overhead, computational cost, and coverage overlap. Section \ref{sec:result_plots} delves into this balance of metrics, showcasing results that validate CU-LVP's effectiveness.

\section{Proposed Method} \label{sec:proposed_method}

\begin{figure}
\centering
\includegraphics[width=0.3\textwidth]{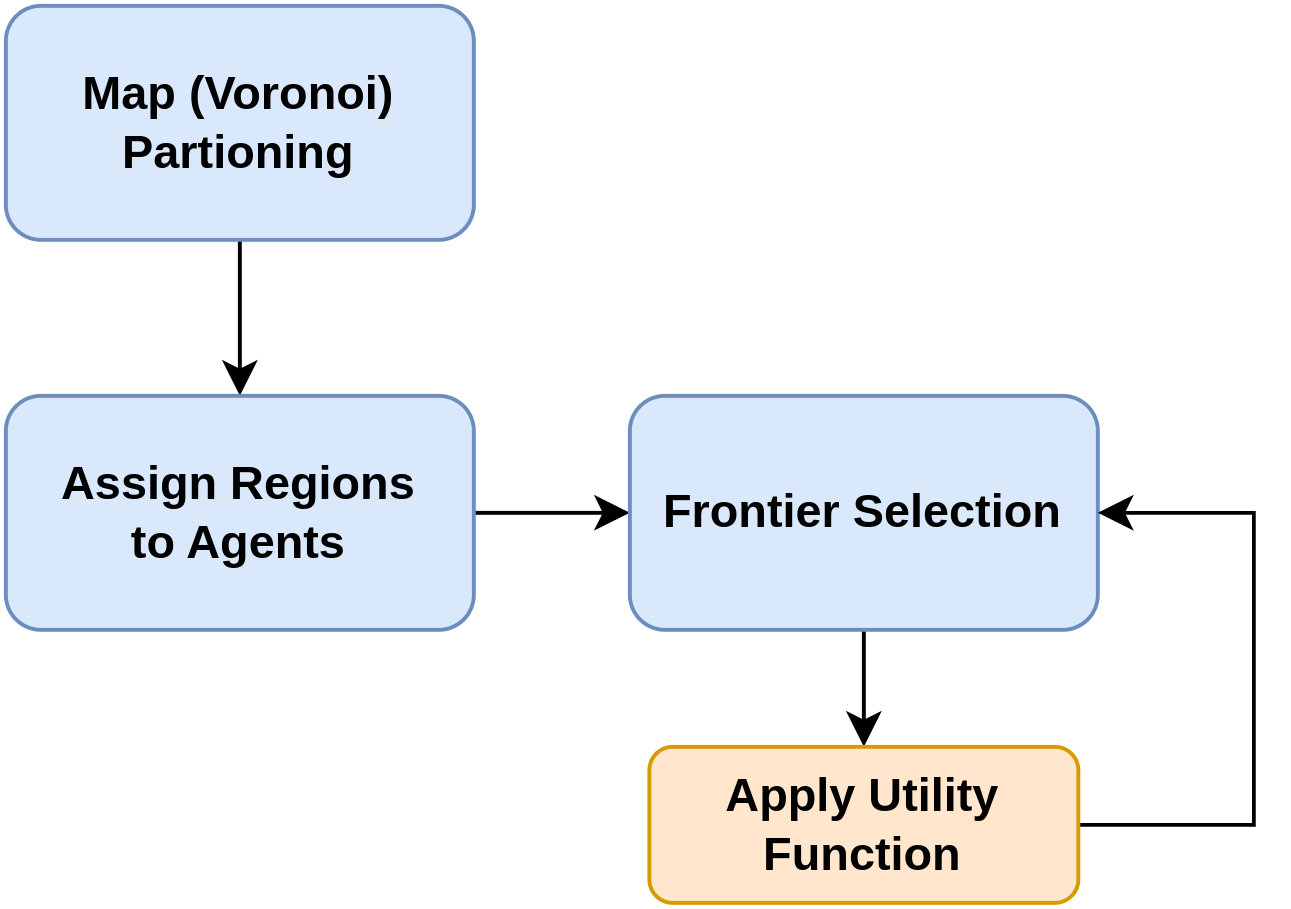}
\caption{Abstract overview of proposed methodology. Implementation details are provideed in Section \ref{sec:proposed_method}.}
\label{fig:prop_approach}
\end{figure}

CU-LVP is a cost-utility approach that enhances frontier selection by combining modifications of previously implemented utility functions. It also effectively addresses the distributed maze problem by selecting frontiers assigned to each agent individually through Voronoi partitioning. Specifically, the process begins with map partitioning, where the maze is divided into subareas using Voronoi diagrams. Each agent is then assigned a specific subarea to explore. The next step involves frontier finding, where the best frontier is selected within each subarea, according to the utility function described in Section \ref{subsec:goal_selection}. The shortest path to each frontier is then calculated, ensuring efficient navigation. Fig. \ref{fig:prop_approach} provides a high-level overview of our methodology, illustrating its application for distributed maze coverage. Before going into more details of the proposed approach it is necessary to define the main entity of our method, \textit{the agent}.

\subsection{Definition of the robotic agent} \label{subsubsec:agent_class}
In our experiments, we define an agent class that represents the behavior of individual agents within a maze environment. The agents are capable of moving in four directions: up, down, left, and right. To avoid collisions, they are prohibited from moving into cells that are already occupied by other agents or obstacles. Each agent keeps a personal map of the areas it has explored in the maze. This map is continuously updated with information collected by other agents within broadcast range in each round. By sharing information about the explored sections of the maze, we simulate typical multiagent communication and behavior for maze exploration. Additionally, the agent's map is updated as it explores new areas within its sensor range, which, by default, extends two blocks in all orthogonal and diagonal directions. This simulates the sensing capabilities of real-world robots, which often have sensors on all sides. Moreover, the agent maintains a list of coordinates corresponding to its assigned Voronoi region earmarked for exploration. During each iteration, the agent employs the utility function described in Section \ref{subsec:goal_selection} to choose a goal exclusively from its unexplored Voronoi coordinates. Once the agent discovers its designated Voronoi coordinate either autonomously or through information exchange with other agents, it redirects its focus to other (unexplored) Voronoi coordinates, thereby continuing its exploration trajectory.

\begin{figure}[htbp]
    \centering
    \includegraphics[width=0.37\textwidth]{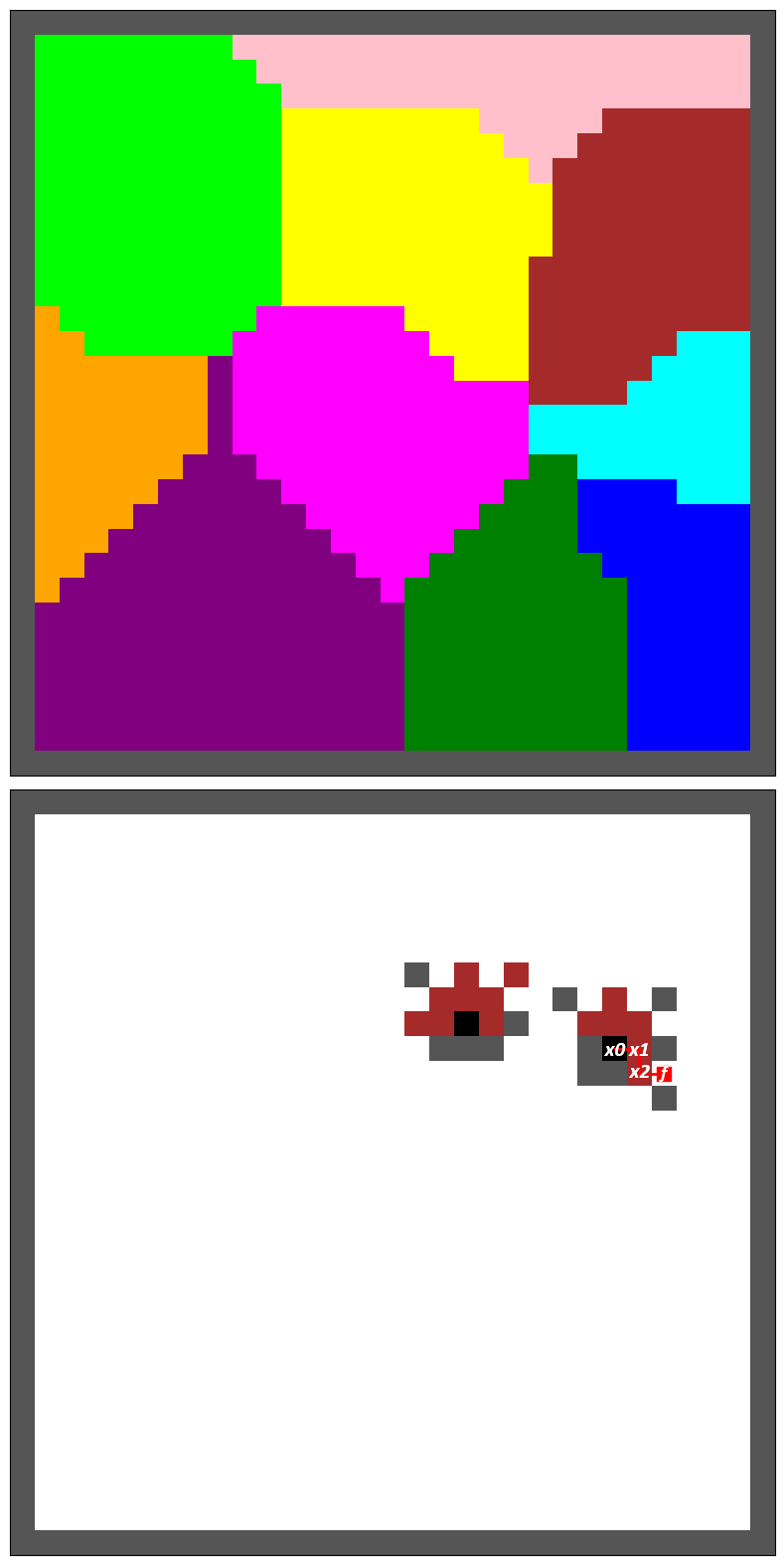}
    \caption{Visualization of the initial Voronoi regions for a randomly 30x30 generated maze (top image). Agents within broadcast range (bottom image) have merged their Voronoi regions and exchanged information about their surroundings. Additionally,
    on the bottom, the red line depicts the wavefront path ($x_1, x_2$) from an agent ($x_0$) to its goal ($f$) within its region. The black cells represent the agents. The colored cells depict the Voronoi regions, with each color representing a different partition, whereas the darker gray cells denote obstacles.}
    \label{fig:agent_view}
\end{figure}

Fig. \ref{fig:agent_view} (in the bottom) shows an agent's explored map at the beginning, following initial information exchange between agents. The map is shown with a two-block view range, as configured in our experiments. If an obstacle is detected, the view beyond it is blocked to mimic real-sensor behavior. The agent class also stores essential parameters for exploration, including the agent's goal (marked with a red dot in Fig. \ref{fig:agent_view}), the next position along the shortest path to the target (indicated by a red line), the current position, and the view range. These parameters allow us to implement realistic agent behavior and exploration scenarios.

\subsection{Voronoi Partitioning}\label{subsec:voronoi_part}
CU-LVP relies on Voronoi partitioning to divide the maze, with each region centered around an agent. Consequently, the number of maze sub-regions in our experiments matches the number of agents. Agents are then tasked with exploring all cells within their assigned region. As they navigate the maze using the utility function described in Section \ref{subsec:goal_selection}, agents within the broadcast range share information with each other, facilitating collaborative exploration. The broadcast range is defined as 25\% of the maximum dimension of the maze. Furthermore, to expedite the exploration process, a mechanism is implemented to merge Voronoi regions when agents come into broadcast range. This merging facilitates a broader scope of exploration for agents, allowing them to explore the maze more effectively.

An example of the Voronoi partition process and broadcast range can be found in Fig. \ref{fig:agent_view}. On the top image,
the starting Voronoi regions are depicted. Initially, all space in the maze is considered as free space, in order for the agents to explore it. The darker gray cells represent the obstacles that the agents find along the way. Additionally, in the bottom image, the Voronoi regions of two agents (depicted as black cells) within the broadcast range have merged. This merging results in the region adopting a red hue, signifying consolidation into a single region. Furthermore, the two agents have exchanged information about their respective surroundings within a sensor range. It's worth noting that the agent's goal (depicted as a red dot in the bottom image) lies within its Voronoi region, signifying correct code execution.

Throughout the exploration process, agents periodically transmit their explored maps centrally, without pulling them. This allows for monitoring of exploration progress and facilitates experiment termination when the maze is explored. Upon completing the exploration of their designated region, agents move to the nearest unexplored Voronoi region, which is decided based on the central map.

Agents also possess the capability to traverse cells in neighboring regions to access their next target within their Voronoi region, overcoming obstacles that might impede progress if they were to confine their movement solely within their designated area. Subsequently, they update their exploration map to reflect their progress. However, the primary focus remains on achieving complete coverage within the assigned region.

\subsection{Goal selection} \label{subsec:goal_selection}
Initially, CU-LVP uses wavefront propagation, starting from the agent's position, to efficiently calculate the shortest path lengths to the frontiers of its assigned region. Among the unexplored cells in the agent's Voronoi partition, the nearest frontier is selected based on distance. If multiple frontiers share the same minimum distance, for each frontier $a_i$ in the nearest frontier list, the utility function described later in this section is computed. The frontier with the highest utility score is chosen as the agent's next target cell.

CU-LVP also ensures that all agents within the broadcast range have different goals by excluding frontiers already assigned to an agent, provided the number of unexplored frontiers is greater than or equal to the number of agents. This differentiation is applied when agents share the same Voronoi region, either due to merging their regions or exploring regions assigned to other agents once they have completed the exploration of their initial regions. By excluding assigned frontiers, agents avoid converging to the same goals, spreading out efficiently across different areas. It is also important to note that this differentiation can only be detected when agents are within broadcast range, enabling them to communicate and share their assigned frontiers, thus avoiding duplication. Additionally, when agents explore the same (or neighboring) regions, one agent may inadvertently explore a goal intended for another. To encounter this, CU-LVP checks at each round whether an agent's goal has been explored by others within its broadcast range. If the goal has been explored, the algorithm recalculates the next optimal goal, enhancing overall exploration efficiency.

The utility function of CU-LVP combines elements from the cost-utility function CU-MNM \cite{cu_mnm_paper} (later referenced as $u_{mnm}$) and the CU-JGR approach \cite{cu_jgr}. Instead of merely considering the neighboring cells (in view range) of the goal cell in the cost-utility function, as CU-JGR does, our method considers the neighboring cells along the whole wavefront path from the current position of the agent to its goal cell (later referenced as $u_{jgr}$). This comprehensive calculation provides a more accurate estimation of the expected explored cells and improves the overall efficiency of the maze exploration task, as demonstrated in the experimental evaluation. Specifically, the formula for this new cost-utility function is as follows:

\begin{equation}
utility(f) = N(u_{mnm}(f)) + \lambda \cdot N(u_{jgr}(f))
\label{eq:utility}
\end{equation}

\noindent where $f$ represents the nearest frontier cell under examination within the agent's Voronoi region, and $N(x)$ denotes the min-max normalization function, ensuring values fall within the range [0, 1].  If cell $x_0$ is the current position of agent $x$, the path $P$ is the sequence $x_1, ..., x_i, ..., x_n$ of cells that the agent has to cross in order to reach the target $f$, which always lies within the agent's assigned region. These cells of path $P$ may not all belong to the region assigned to this agent, allowing the agent to overcome obstacles that might impede progress if it were confined solely within its designated area. As the agent follows path $P$, several unexplored cells within its view range will be explored, enhancing the overall exploration efficiency if the agent selects the specific frontier $f$.

\begin{figure}[htbp]
    \centering
    \includegraphics[width=0.40\textwidth]{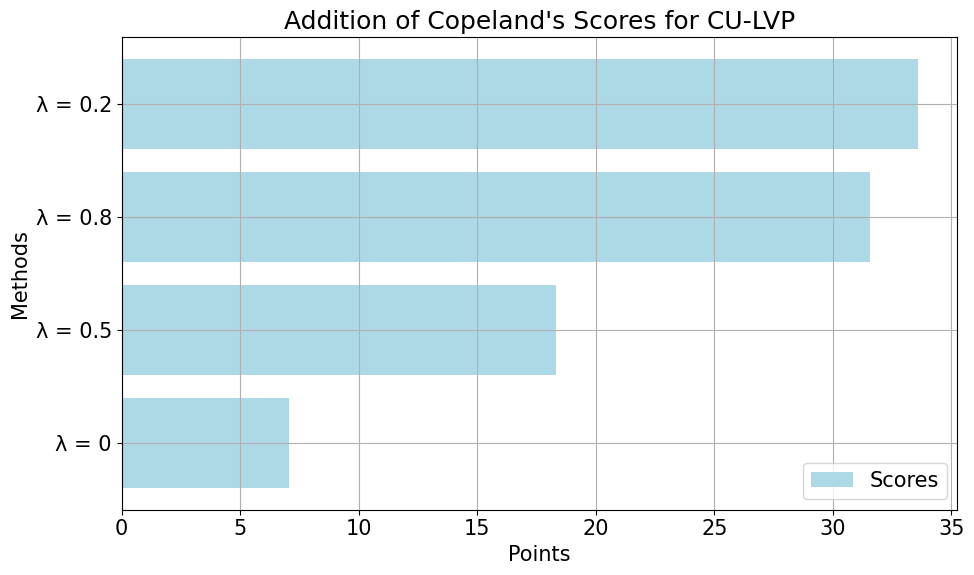}
    \caption{Copeland Comparison for different $\lambda$ values of CU-LVP across 100 experiments in various maze sizes (namely 15x15, 30x30, and 50x50).}
    \label{fig:new_cu_lambda}
\end{figure}

Moreover, the method incorporates a parameter $\lambda$, which was determined to be optimal at 0.2 after conducting 100 experiments across maze dimensions of 15x15, 30x30, and 50x50. A more comprehensive comparison of the outcomes across different $\lambda$ values is shown in Fig. \ref{fig:new_cu_lambda}, using Copeland's method. Copeland's method \cite{copeland3, copeland1, copeland2} is a voting technique, which allows ranking multiple methods that are compared on a set of experiments with varying parameters (e.g. maze sizes, number of agents or partitions, etc.) to provide a unified ranking. The technique involves pairwise comparisons of the average scores obtained across the experiments for each $\lambda$ value. Based on the results, we can determine which $\lambda$ value consistently received more votes, thereby establishing its superiority across the experiments.

The first utility score of the utility function (\ref{eq:utility}), $u_{mnm}$, for the (nearest) frontier cell $f$ is defined as:

\begin{equation}
u_{mnm}(f) = \sum_{i=1}^{n} dist[(X_{f}, Y_{f}), (X_{r_i}, Y_{r_i})]
\end{equation}

\noindent where, $x_0=(X_{r_i}, Y_{r_i})$ represents the position of agent $i$, $x_n=(X_{f}, Y_{f})$ denotes the position of frontier $f$ within its Voronoi region, and $n$ signifies the number of agents. Also, $u_{jgr}$ for the frontier cell $f$ is defined as:

\begin{equation}
u_{jgr}(f) = \sum_{x_i \in P} Unex(x_i, r)
\end{equation}

\noindent where $Unex(x, r)$ represents the number of unexplored cells within the agent's view range at cell $x$. $x_i$ denotes the individual elements (cells) of the $path$, designated as the wavefront path, guiding the agent towards the target frontier $f$. Once again, the frontier(s) $f$ considered are within the agent's currently assigned Voronoi region.

CU-LVP also incorporates collision avoidance strategies to navigate around other agents or obstacles. Specifically, if an agent is unable to reach the nearest frontier due to obstacles or blocked paths caused by other agents within broadcast range, it remains stationary until the obstructing agents move. However, this is a rare occurrence due to the integration of Voronoi partitioning. As detailed in Section \ref{subsec:voronoi_part}, agents prioritize exploring their own Voronoi regions, which effectively prevents collisions throughout the experiments. Consequently, collisions may occur only when two or more agents navigate through the same Voronoi region.

\section{Evaluation Framework} \label{sec:experimental_eval}

To evaluate the performance of the newly proposed CU-LVP and compare it with other state-of-the-art techniques that we adapted to the distributed maze problem, we conducted experiments using randomly generated mazes and a varying density of obstacles. Several metrics were used to evaluate the performance of different methods under various conditions. The experiments ran 500 times and the average scores are reported. The code and results of our experiments are available in a code repository (\href{https://github.com/manouslinard/multiagent-exploration/}{https://github.com/manouslinard/multiagent-exploration/}).

\subsection{Maze Generation}\label{subsubsec:maze_gen}

For generating mazes, we adapted the methodology described in \cite{mazegen}. This method involves moving an agent randomly in orthogonal directions (up, down, left, and right) within a grid to construct the maze. Also, we extended this approach by introducing a probability factor for each obstacle cell, determining whether an obstacle should remain or be converted to empty space, thus allowing for the creation of mazes with varying levels of complexity. The function also permits the specification of maze dimensions.

The maze is represented as a 2D array with four cell states: unexplored cells (-1), free space (0), obstacles (1), and agent cells (2). When an agent views its surroundings, cells within its range are marked using 0 for free space and 1 for obstacles.

\begin{figure}[htbp]
    \centering
    \includegraphics[width=0.40\textwidth]{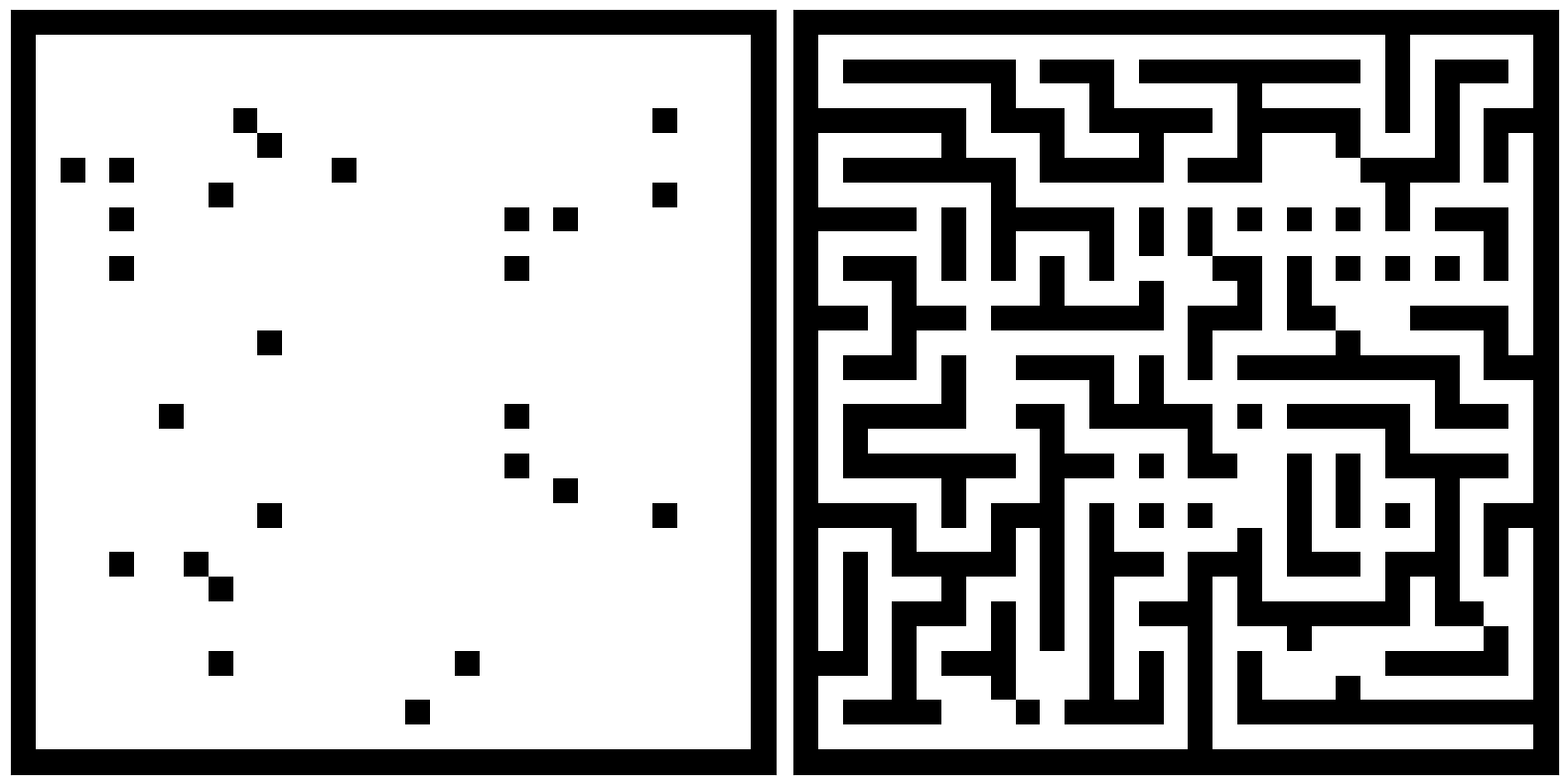}
    \caption{Randomly generated 30x30 mazes from the experiments. The left maze has an obstacle probability of 15\%, whereas the right maze has an obstacle probability of 85\%.}
    \label{fig:mazes}
\end{figure}

For our experiments, we generated 30x30 mazes with varying obstacle probabilities to create environments of differing complexities, set at 85\% and 15\%. This range enabled a thorough evaluation of the algorithms' performance under different conditions. An obstacle probability of 85\% simulates a complex, maze-like stage, whereas 15\% resembles a sparsely obstructed area. To ensure robustness in our evaluation, new mazes were generated for each experiment, providing fresh scenarios to test the efficiency of the algorithms. Fig. \ref{fig:mazes} shows two randomly generated 30x30 mazes, one with an obstacle probability of 15\% and the other with 85\%, demonstrating the diversity in maze complexity examined in our experiments.

\subsection{Baseline methods}\label{subsec:baseline_methods}

The baseline methods used to compare CU-LVP are described in Table \ref{tab:baseline_table}. Each method has been implemented from scratch based on the corresponding papers and adapted to the distributed maze problem using Voronoi partitioning. The specifics of the modifications for adapting these methods to partitioned mazes are detailed in Section \ref{subsubsec:vor_baseline}, whereas parameter tuning is discussed in Section \ref{subsubsec:base_tuning}.

\begin{table}
    \caption{Methods used for comparative analysis in distributed maze exploration.}
    \centering
    \begin{tabular}{|p{0.2\linewidth}|p{0.7\linewidth}|}
    \hline
    \multicolumn{1}{|c|}{\textbf{Name}} & \multicolumn{1}{c|}{\textbf{Description}} \\
    \hline
    HEDAC & Utilizes potential fields to explore the maze \cite{hedac_maze}. \\
    \hline
    Nearest-Front & Selects the closest frontier to the agent \cite{yamauchi1998frontier}. \\
    \hline
    CU-BSO & Uses wavefront propagation \& selects the frontier with less robots near it \cite{cu_bso}. \\
    \hline
    CU-MNM & Cost-utility method considering the distances of frontiers from all agents \cite{cu_mnm_paper}. \\
    \hline
    CU-JGR & Cost-utility method estimating the expected explored cells in the target cell \cite{cu_jgr}. \\
    \hline
    \end{tabular}
    \label{tab:baseline_table}
\end{table}

\subsubsection{Voronoi Partitioning for Baseline Methods} \label{subsubsec:vor_baseline}
The baseline methods were adapted to the distributed maze problem. Specifically, frontier exploration methods like CU-MNM, CU-JGR, CU-BSO, and Nearest Frontier follow the same principles for Voronoi integration as CU-LVP, as discussed in Section \ref{subsec:voronoi_part}. This is due to the fact that they are similar to CU-LVP, since CU-LVP is also a frontier exploration method. 

HEDAC on the other hand differs in its adaptation for the distributed maze problem. Specifically, HEDAC constrains the agent's movements within its designated Voronoi region, similar to HYC \cite{zheng2022distributed}. This distinction arises from the characteristic of temperature/potential field methods, which select one neighboring cell at a time. To ensure navigation solely within the agent's Voronoi region, we consider only cells within that region, treating all other unexplored cells outside of it as obstacles for accurate potential field calculations.

However, this approach may encounter challenges in maze environments, where obstacles hinder progress, potentially causing the agent to become trapped and unable to explore its entire Voronoi region. To mitigate this, we refined the HEDAC algorithm. Once the agent has explored all reachable cells, we enhance the attraction force along the wavefront path, guiding the robot toward the nearest frontier. Consequently, the agent can navigate situations where its Voronoi region becomes inaccessible due to limited movement within it and the presence of obstacles, effectively addressing the distributed maze problem.

\subsubsection{Parameter Tuning for Baseline Methods} \label{subsubsec:base_tuning}
Certain baseline methods require specific parameter settings to achieve optimal performance. For example, in HEDAC, the ``iterations" parameter, as described in the original paper \cite{hedac_maze}, determines how many times the attractive force should be recalculated to produce the final attractive force for each cell. Through experimentation in distributed mazes, we found that setting this parameter to 100 yielded the best results. Lower values caused agents to become trapped in local minima, impeding exploration progress, whereas higher values significantly increased computation time. Additionally, HEDAC includes a parameter denoted $a$, which we found to perform optimally when set to 10. Furthermore, the anti-collision (AC) condition outlined in the original HEDAC paper is set to ON in our experiments. This means that if an agent has information that one of the other agents is currently standing on one of its neighboring nodes, the agent does not consider this node for its next position. We also extended the original HEDAC approach by setting the agent's view range to two blocks.

\begin{figure}
  \centering
  \includegraphics[width=0.47\textwidth]{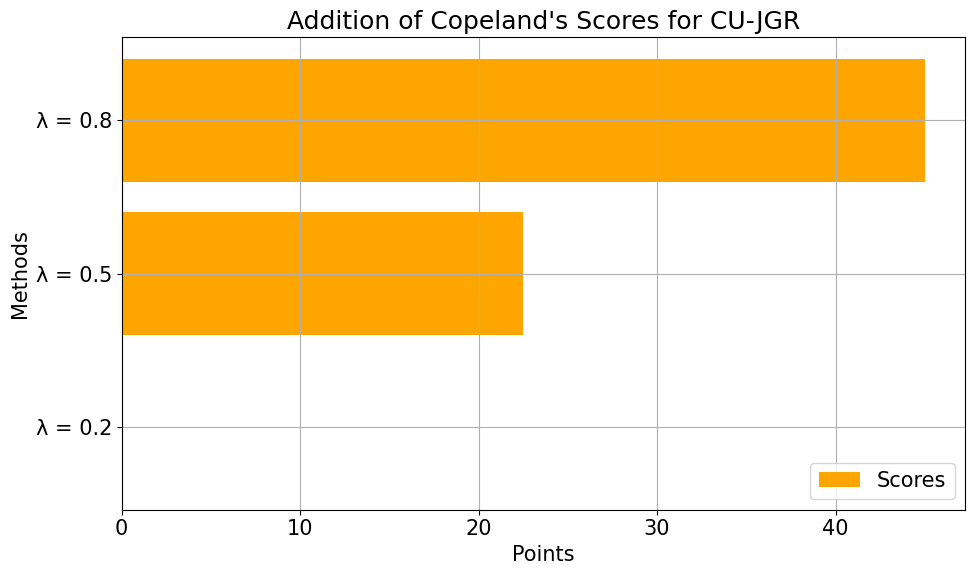}
  \caption{Copeland Comparison for different $\lambda_{jgr}$ values of CU-JGR across 100 experiments in various maze sizes (namely 15x15, 30x30, and 50x50). }
  \label{fig:lambda_comp}
\end{figure}

The CU-JGR algorithm also requires tuning of the parameter $\lambda_{jgr}$. Through experimentation in partitioned mazes, we determined that setting $\lambda_{jgr}$ to 0.8 resulted in more efficient maze exploration. Fig. \ref{fig:lambda_comp} shows the results obtained for different $\lambda_{jgr}$ values, using Copeland's method.

\subsection{Evaluation metrics}\label{subsec:experiment_metrics}
To compare the methods, we utilized evaluation metrics proposed by Yan \textit{et al.} \cite{yan2015metrics}. Additionally, we tracked the number of exploration rounds to measure the repetition of the exploration process. A metric for estimating the communication cost between agents was also developed. The evaluation metrics used in this study include:
i) Exploration Time, ii) Exploration Rounds, iii) Exploration Cost, iv) Exploration Efficiency, v) Map Quality, and vi) Communication Cost.

\subsubsection{Exploration Time}
\label{subsubsec:exploration_time}
The exploration time metric measures the total time required for a robot fleet to complete an exploration mission. The timer starts when at least one robot begins exploring and stops when the fleet collectively achieves 100\% terrain exploration. This metric is measured in wall-clock time, reflecting the duration in days, hours, minutes, and seconds spent on the exploration task.

In our experiments, exploration time measures the computational time for each method, as the movement time is instantaneous. This is because the agents move within a simulated grid (a 2D array), where we simply change the cell value from 0 (free space) to 2 (occupied by an agent) to signify agent movement. Specifically, the (exploration) time is calculated by finding the average time of each round and multiplying it by the number of rounds. Since the agents operate sequentially (one agent after the other performs a step in each round), we calculate the average round time by dividing the time needed for all agents to make a step to the number of agents:

\begin{equation}
explorationTime(n) = \frac{\sum_{i=1}^{n} t_{s_i}}{n} \cdot R
\label{eq:exp_time}
\end{equation}
\noindent where $t_{s_i}$ is the time needed for an agent to make a step.

\subsubsection{Exploration Rounds}\label{subsubsec:exploration_rounds}
Exploration rounds ($R$) represent the iterative cycles of the exploration process, including both information exchange among agents and the algorithm's repetition until maze completion. These iterations measure the number of times the agents have communicated with each other, as information sharing occurs in each round for agents within broadcast range. Exploration iterations conclude when the entire maze has been fully explored.

\subsubsection{Exploration Cost}
\label{subsubsec:exploration_cost}
The exploration cost is based on user specifications, including factors such as energy consumption by computational resources (e.g., CPU, RAM, and network bandwidth) and robot-related expenses like acquisition, handling, and maintenance costs. Their definition of the exploration cost metric requires summing the distances traveled by each robot during the collaborative maze exploration. This metric can also be utilized to measure repetitive coverage, as it is calculated using the following formula:

\begin{equation}
explorationCost(n) = \sum_{i=1}^{n} d_i
\label{eq:exp_cost}
\end{equation}

\noindent where $n$ is the number of robots in the fleet, and $d_i$ is the distance traveled by robot $i$.

\subsubsection{Exploration Efficiency}
\label{subsubsec:exploration_efficiency}
Exploration efficiency relates to the amount of environmental information acquired, inversely proportional to the costs incurred by the robot fleet. It is described by the following formula:

\begin{equation}
explorationEfficiency(n) = \frac{M}{explorationCost(n)}
\label{eq:exp_eff}
\end{equation}
\noindent where $n$ is the number of robots in the fleet and $M$ is the total explored cells.

\subsubsection{Map Quality}
\label{subsubsec:map_quality}
The map quality is defined as the overlap of the explored map and the ground truth map as a ratio to the total area of the ground truth map $P$:

\begin{equation}
mapQuality = \frac{M - A(mapError)}{P}
\label{eq:map_quality}
\end{equation}

\noindent where $M$ are the total explored cells, $P$ is the total area of the ground truth map, and $A(mapError)$ is the area occupied by error cells. Error cells are those in the explored map that differ in value from the corresponding cells in the ground truth map. In our experiments, the $mapQuality$ metric consistently reaches 100\% due to the efficiency of the algorithms and the design of the mazes, which facilitate agents' access to all areas for exploration.

\subsubsection{Communication Cost} \label{subsubsec:comm_cost}
The communication cost refers to the cells transferred from one agent to others within range during communication intervals. During each (exploration) round, an agent shares its entire explored stage with other agents within its communication range. Considering the four possible cell states (-1, 0, 1, and 2), we require 2 bits to represent each cell. Additionally, since each communication exchange occurs bidirectionally between two agents, the equation for calculating the communication cost is formulated as follows:
\begin{equation}
commCost = 4 \cdot rows \cdot columns
\label{eq:comm_cost}
\end{equation}
Here, the factor of 4 accounts for the 2 bits needed per cell multiplied by 2, representing the number of communicating agents. The variables $rows$ and $columns$ denote the dimensions of the maze, as the agent transmits its entire explored stage during each communication cycle. For instance, in a 30x30 maze, both the number of rows and columns is 30.

\subsection{Hardware description} \label{subsec:hardware_desc}
The final experiments were conducted on a PC running Linux equipped with an Intel(R) Core(TM) i5-1235U CPU. To optimize computational efficiency, the Python multiprocessing library was employed, utilizing all available processing cores. Detailed results of these experiments are presented in Section \ref{sec:result_plots}.

\section{Experimental Results} \label{sec:result_plots}
The results consist of 500 experiment repetitions conducted with varying numbers of agents; specifically 1, 2, 4, 6, 8, 10, 15, and 20. Each agent configuration explored mazes of dimensions 30x30 with obstacle probabilities set at 85\% and 15\%. These experiments included both baseline methods and the newly proposed CU-LVP, as depicted in the resulting plots. Fig. \ref{fig:res_all} (refer to Section \ref{subsec:experiment_metrics} for details on the metrics) shows the average scores and standard deviations obtained from the experiments for each agent group and exploration method.

For the final evaluation, we employed the Composite Index Scoring method \cite{chen2022composite, chakrabartty2017composite}. This method is a powerful and popular tool for providing an overall measure of a subject (in our case, a method) by summarizing a group of measurements (component indices) of different aspects of the subject/method. It is widely used in economics, finance, policy evaluation, performance ranking, and many other fields. According to \cite{chen2022composite}, the most widely used approach is to use a linear combination of the component indices with specified weights. In our case, the components are the metrics defined in Section \ref{subsec:experiment_metrics}, and the weights are all set to 1, as all metrics are equally important.

Specifically, we produced the Composite Index Scoring plot by normalizing all the metrics for each agent group using min-max scaling. This means, for example, that the maximum value of the min-max scaling for a setup with N agents is the largest (average) metric value (for the 500 repetitions) for the N-agents' setup only (and similarly for the minimum value). We ranked each method in each agent group for all metrics based on these normalized values. Then, we summed these rankings using equal weights to create the final Composite Index Scoring plot. The resulting Composite Index plot is displayed in Fig. \ref{fig:comp_index_plot} and detailed in Section \ref{subsec:comp_index_plot}. Greater values in the Composite Index Scoring denote better method performance.

\subsection{Results of Metrics} \label{subsec:res_all_metrics}
\begin{figure}
\centering
\includegraphics[width=0.47\textwidth]{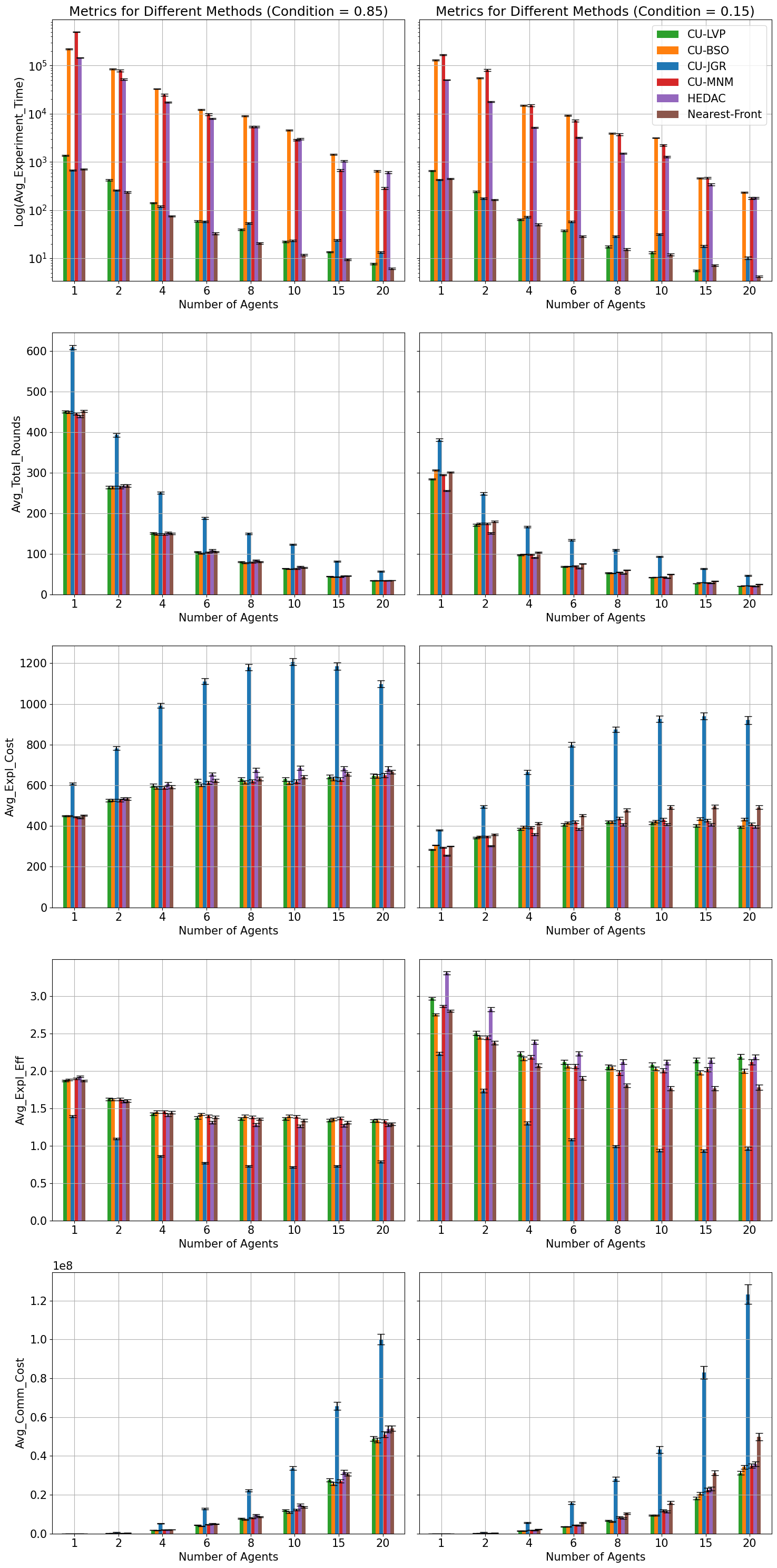}
\caption{Results of 500 experiments, utilizing the metrics described in Section \ref{subsec:experiment_metrics} for CU-LVP and baseline methods.}
\label{fig:res_all}
\end{figure}

The results of the metrics showcase the overall efficiency of CU-LVP. The first row of plots in Fig. \ref{fig:res_all} displays the Exploration Time results on a logarithmic scale for better visibility. CU-LVP is among the least computationally expensive methods across both maze topologies, comparable to Nearest-Frontier, i.e. the simplest baseline method. CU-JGR, another computationally efficient method, performs similarly to CU-LVP in complex mazes (85\% obstacles). However, in simpler mazes (15\% obstacles) and with more than 4 agents, CU-LVP proves to be more efficient. Other methods like CU-MNM, CU-BSO, and HEDAC are more computationally expensive, leading to longer exploration times. This efficiency is attributed to CU-LVP's use of wavefront propagation to find the nearest frontiers, resulting in faster utility function calculations and overall algorithm performance.

The second row of Fig. \ref{fig:res_all} shows the Exploration Rounds Metric. In mazes with dense (i.e. 85\%) obstacles, CU-LVP performs comparably to other methods except for CU-JGR, which has the highest number of exploration rounds. In mazes with few (i.e. 15\%) obstacles, CU-LVP generally ranks second to HEDAC when there are fewer agents. As the number of agents increases, performance across methods becomes similar, except for CU-JGR, which consistently performs the worst, exhibiting the highest exploration rounds in both maze scenarios.

The third row presents the Exploration Cost results, which can also be interpreted as repetitive coverage results. CU-JGR again performs the worst in both scenarios. Similarly, HEDAC falls short compared to other methods in the dense obstacle setting, gradually ranking as the second-worst in exploration cost as the number of agents increases. CU-LVP remains competitive in complex mazes, improving its performance as the number of agents increases. In simpler mazes, CU-LVP typically ranks second to HEDAC with fewer agents, but this disparity diminishes as the number of agents rises, eventually achieving parity with HEDAC.

The fourth row displays Exploration Efficiency performance. CU-JGR has the lowest efficiency in both maze conditions, followed by HEDAC in 85\% obstacle probability mazes. CU-LVP's efficiency is comparable to other cost-utility methods in dense mazes. In simpler mazes, HEDAC is the most efficient, followed by CU-LVP, with both methods performing equally well as the number of agents increases.

The fifth and final row shows Communication Cost results. CU-JGR consistently has the highest communication cost in all maze topologies. Communication costs increase for all methods as the number of agents rises. CU-LVP maintains one of the lowest communication costs overall, especially in simpler mazes where it consistently ranks first. This is likely due to the effective dispersion of robots by CU-LVP, reducing communication overhead.

\subsection{Composite Index Scoring} \label{subsec:comp_index_plot}
\begin{figure}[t]
\centering
\includegraphics[width=0.45\textwidth]{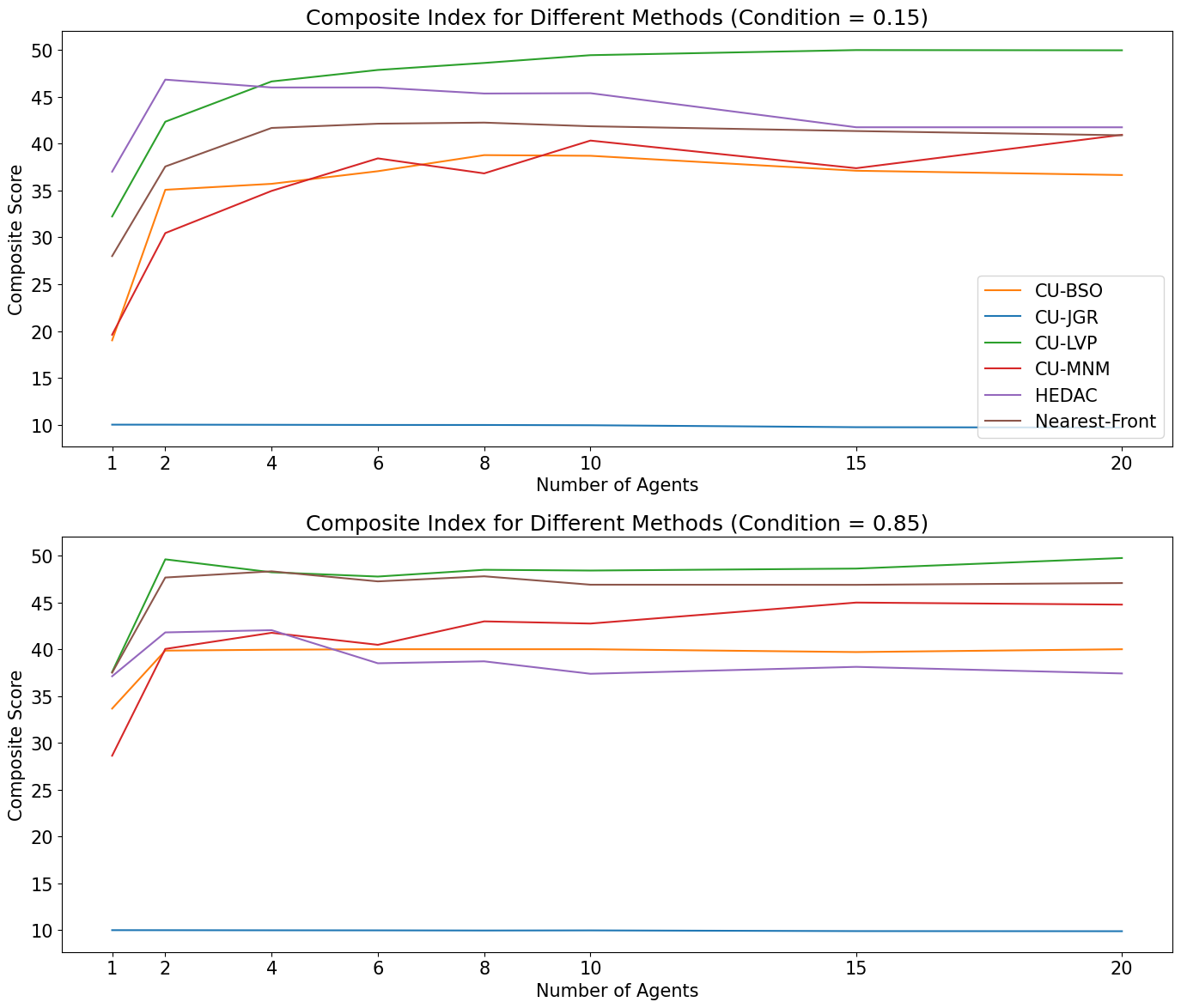}
\caption{Composite Index Scores across different methods.}
\label{fig:comp_index_plot}
\end{figure}

Examining the Composite Index depicted in Fig. \ref{fig:comp_index_plot}, it becomes evident that CU-LVP achieved the highest overall score in both maze topologies. Particularly, it outperforms all other methods in complex mazes (with 85\% obstacle probability) and excels when more than 4 agents are used in the exploration of simpler mazes. However, for 1 and 2 agents in less complicated mazes, HEDAC exhibits superior performance. The advancement of CU-LVP over other baselines stems from its competitive results across all maze settings in metrics such as Exploration Rounds, Exploration Cost, Communication, and Efficiency, while maintaining significantly lower computational complexity compared to other baselines, except Nearest Frontier, i.e. the simplest frontier exploration method.

\section{Conclusions} \label{sec:conclusions}
This work introduced CU-LVP, a novel method for multiagent maze exploration in distributed mazes, which demonstrates competitive results across various metrics. Notably, CU-LVP exhibits exceptional performance in Exploration Time (computational complexity), showcasing its efficiency by requiring minimal computations to explore mazes effectively. The effectiveness of CU-LVP is further evidenced in the Composite Index plot, where it emerges as the top performer across both complicated and simpler mazes, striking a balance between swift exploration, low computational overhead, minimized communication between agents, and reduced coverage overlap.

Moving forward, our research endeavors will focus on integrating successful elements from other methodologies, such as HEDAC, into the utility function to enhance performance in simpler mazes with fewer agents. Additionally, we plan to incorporate cutting-edge techniques in the baseline methods, such as the recently introduced reinforcement learning method CQlite \cite{latif2024communication}, which shares similarities with CU-LVP in emphasizing frontier exploration in distributed mazes. Moreover, the application of these methods in real-world settings holds promise for providing invaluable insights into their performance and efficiency when confronted with practical complexities. Such insights will fuel further advancements and adaptations, enabling broader applicability. 

\section*{Acknowledgment}
The research leading to these results has received funding from the European Union’s Horizon Europe research and innovation programme under grant agreement No 101073876 (Ceasefire). This publication reflects only the authors views. The European Union is not liable for any use that may be made of the information contained therein.

\bibliographystyle{IEEEtran}
\bibliography{main}

\phantomsection
\begin{IEEEbiography}[{\includegraphics[width=1in,height=1.25in,clip,keepaspectratio]{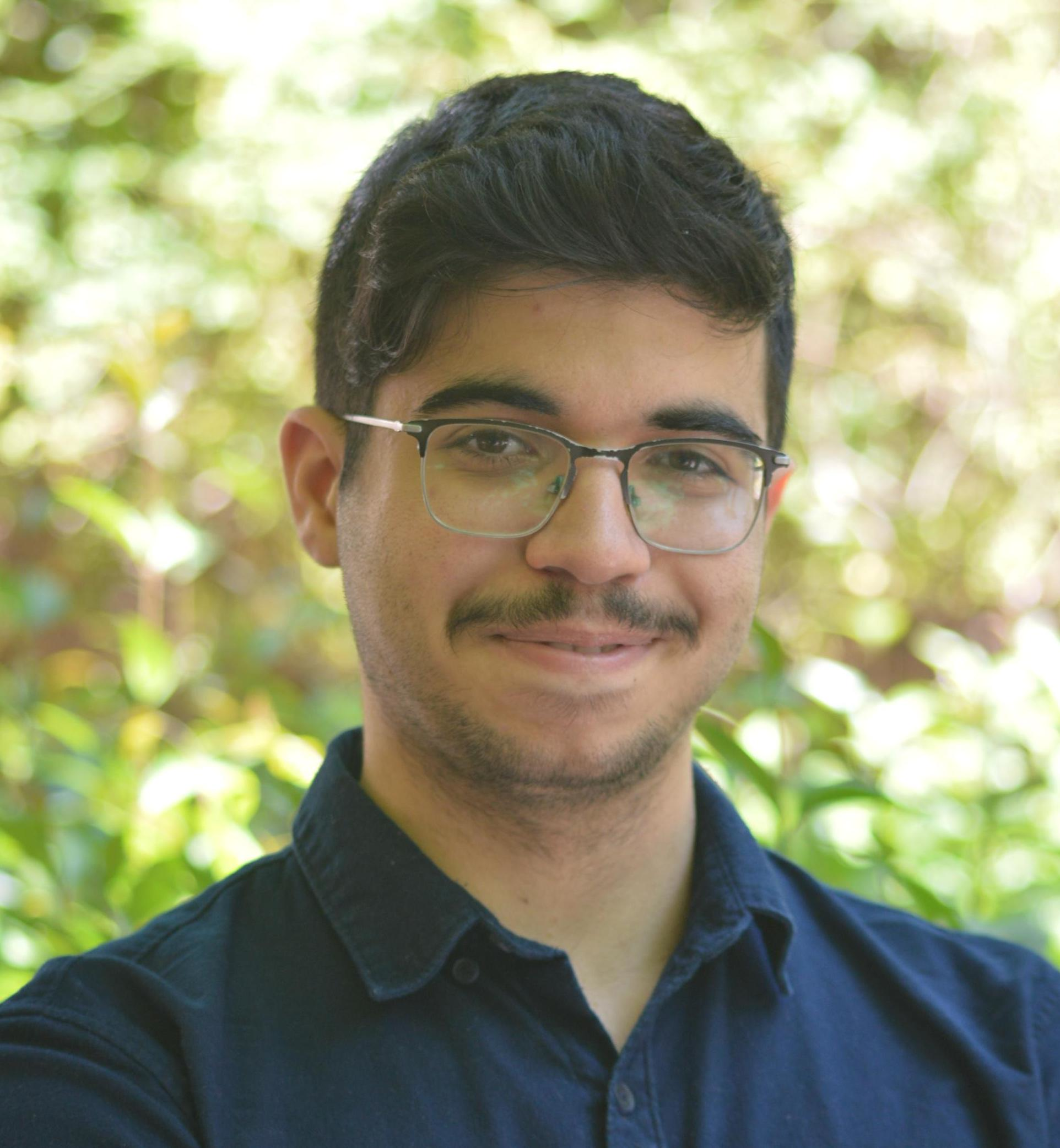}}]{Manousos Linardakis}
is currently an undergraduate student at the Department of Informatics and Telematics, Harokopio University of Athens. He has conducted research in multirobot exploration and hand gesture recognition and has experience in managing laboratory systems, implementing DevOps practices, and developing decision support software. He also has participated in the Google Summer of Code as an open-source developer. His research interests include data science, machine learning, robotics, and multiagent systems.
\end{IEEEbiography}

\begin{IEEEbiography}[{\includegraphics[width=1in,height=1.25in,clip,keepaspectratio]{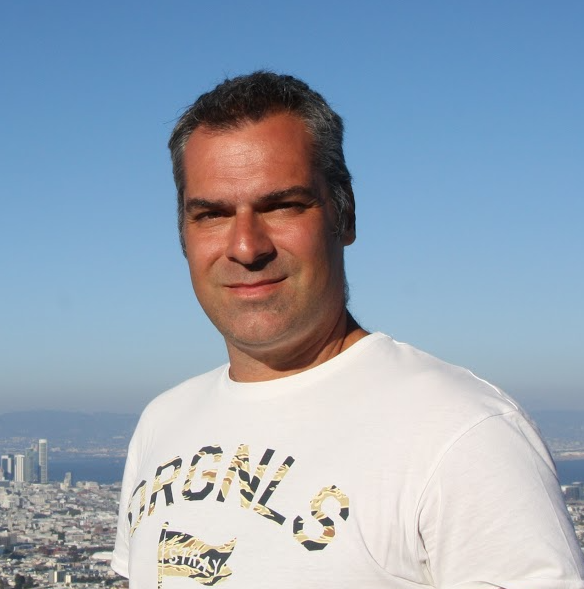}}]{Iraklis Varlamis}
(Member, IEEE) received the M.Sc. degree in information systems engineering from the University of Manchester Institute of Science and Technology, Manchester, U.K., and the Ph.D. degree from the Athens University of Economics and Business, Athens, Greece. He is currently a Professor of data management with the Department of Informatics and Telematics, Harokopio University of Athens (HUA), Kallithea, Greece. He has more than 250 articles published in international journals and conferences and more than 5000 citations on his work. He holds a patent from the Greek Patent Office for a system that thematically groups web documents using content and links. His research interests include data mining and social network analytics to recommender systems for social media and real-world applications. He is the Scientific Coordinator for HUA in several EU (H2020, ECSEL, REC) and Qatar (QNFR) projects as well as in national projects.
\end{IEEEbiography}

\begin{IEEEbiography}[{\includegraphics[width=1in,height=1.25in,clip,keepaspectratio]{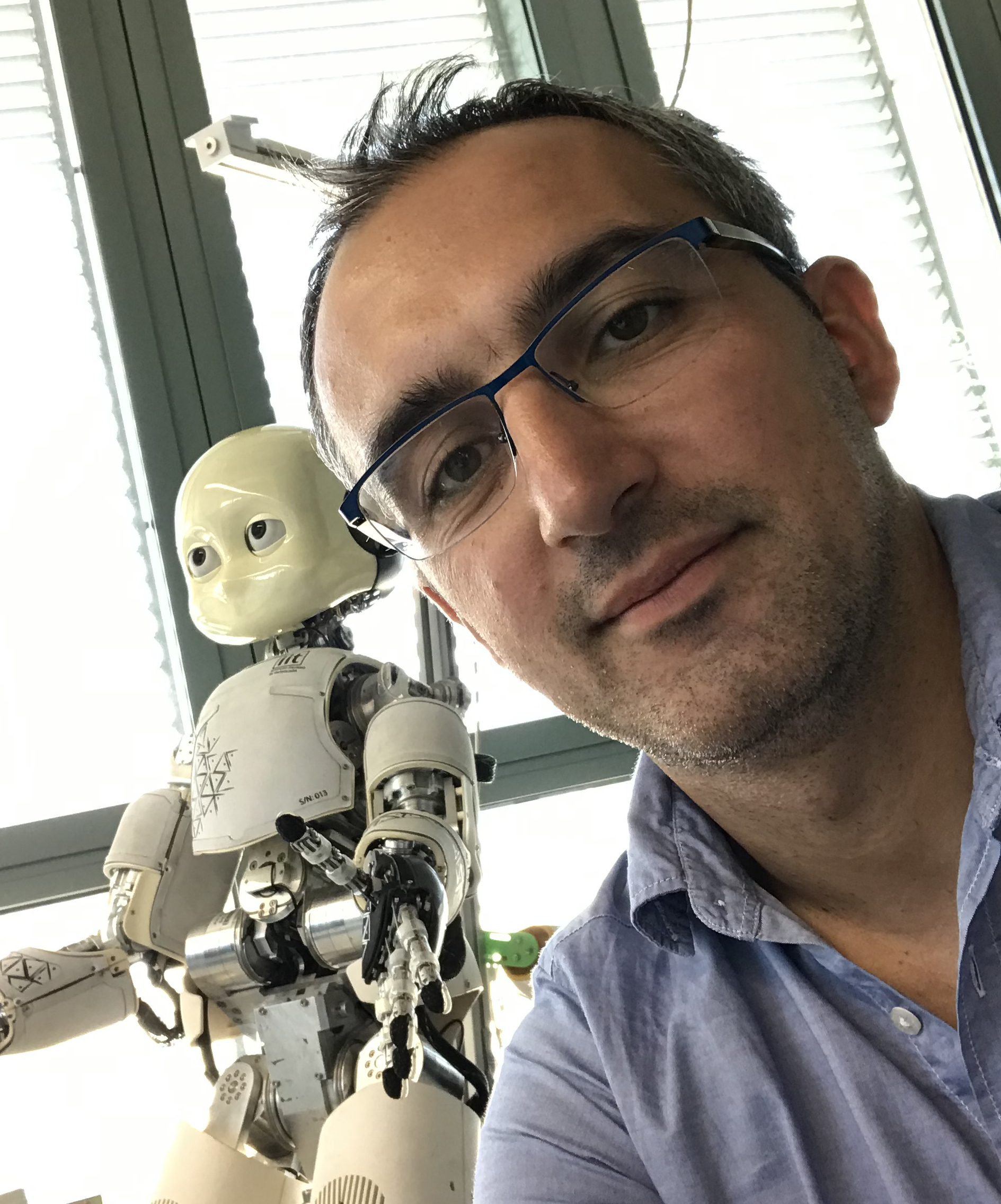}}]{Georgios Th. Papadopoulos}
(M) is an Assistant Professor in the area of Computer Graphics and Computational Vision at the Department of Informatics and Telematics of the Harokopio University of Athens in Greece. He received the Diploma and Ph.D. degrees in electrical and computer engineering from the Aristotle University of Thessaloniki (AUTH), Thessaloniki, Greece. He has worked as a Post-doctoral Researcher at the Foundation For Research And Technology Hellas / Institute of Computer Science (FORTH/ICS) and the Centre for Research and Technology Hellas / Information Technologies Institute (CERTH/ITI). He has published over 70 peer-reviewed research articles in international journals and conference proceedings. His research interests include computer vision, artificial intelligence, machine/deep learning, human action recognition, human-computer interaction and explainable artificial intelligence. Dr. Papadopoulos is a member of the IEEE and the Technical Chamber of Greece.
\end{IEEEbiography}
\end{document}